\documentclass{article}
\usepackage{fleqn}
\usepackage{epsfig}
\usepackage{amsmath,amssymb}
\begin{document}

\begin{center}
{\bf\Large Nonminimal Macroscopic Models of a Scalar Field Based on Microscopic Dynamics. I. Extension of the Theory for Negative Masses.}\\[12pt]
Yu.G. Ignat'ev\\
Kazan Federal University,\\ Kremlyovskaya str., 35,
Kazan 420008, Russia
\end{center}

{\bf keywords}: Relativistic Kinetics, Phantom Scalar Fields,
Scalar Interaction of Particles, Negatives Masses.

\begin{abstract}
The article proposes generalizations of the macroscopic model of plasma of scalar charged particles to the cases of inter-particle interaction with multiple scalar fields and negative effective masses of these particles. The model is based on the microscopic dynamics of a particle at presence of scalar fields. The theory is managed to be generalized naturally having strictly reviewed a series of its key positions depending on the sign of particle masses. Thereby, it is possible to remove the artificial restriction contradicting the more fundamental principle of action functional additivity.
\end{abstract}
\newcommand{\beqdis}[2]{\begin{equation}\label{#1}
{\displaystyle #2} \end{equation}}
\newcommand{\dsp}{\displaystyle}
\newcommand{\Req}[1]{(\ref{#1})}
\newcommand{\krugskob}[1]{\left(#1\right)}
\newcommand{\kvadrskob}[1]{\left[#1\right]}
\newcommand{\figurskob}[1]{\left\{#1\right\}}
\newcommand{\Pg}{{\rm P}}
\newcommand{\PAcontr}[2]{\Pg^{#1} - e_{#2}A^{#1}}
\newcommand{\Lee}[1]{\stackunder{#1}{\rm L}}
\newcommand{\dx}[1]{\partial_{#1}}
\newcommand{\qf}[1]{\displaystyle 1 + \frac{q_{#1}\Phi}{m_{#1}}}
\newcommand{\Eps}{{\cal E}}
\newcommand{\Const}{\mathop{\rm Const}\nolimits}
\newcommand{\Ham}{{\cal H}}
\newcommand{\dfeq}{\stackrel{def}{=}}
\newcommand{\xsort}[2]{\stackunder{#1}{#2}}
\newcommand{\onetwo}{\frac{1}{2}}
\newcommand{\dual}[1]{\stackrel{*}{#1}}
\newcommand{\anti}[1]{\overline{#1} \,}
\newcommand{\dpox}[2]{\displaystyle \frac{\partial #1}{\partial
#2}}
\newcommand{\xsorancov}[3]{\xsort{#1}{\anti{#2}}\,_{#3}}
\newcommand{\intx}[1]{\displaystyle \int\limits_{#1}^{}}
\newcommand{\sumx}[1]{\displaystyle \sum\limits_{#1}^{}}
\newcommand{\Puass}[1]{\left[\Ham\,, #1\right]}
\newcommand{\llsim}{\stackrel{<}{\sim}}
\newcommand{\noneq}[1]{\begin{equation} #1 \nonumber\end{equation}}
\def\stackunder#1#2{\mathrel{\mathop{#2}\limits_{#1}}}

This work was founded by the subsidy allocated to Kazan Federal University for the state assignment in the sphere of scientific activities.

\section{Introduction}
\label{intro}

In the previous articles the Author considered the sta\-ti\-sti\-cal systems of
scalar charged particles and constructed the cosmological models based on such systems \cite{Ignatev1,Ignatev2,Ignatev3,Ignatev4}.
Particularly, in \cite{Ignatev2,Ignatev3} it was obtained the self-consistent set of equations de\-scri\-bing a statistical system of particles
 with scalar interaction. In the recent Author's works \cite{YuNewScalar1,YuNewScalar2,YuNewScalar3}\footnote{see also monographes \cite{Yubook1,Yubook2}.} the macroscopic theory of the statistical systems has been improved significantly in part of formalism development and also it has been extended to the case of fantom scalar fields\footnote{see also \cite{Ignatev14_1} and review \cite{Yu_stfi14}.}. In \cite{Ignatev14_1,Ignat14_2} a question about possibility to include negative effective particle masses in the theory has been raised. The negative answer to this question has been given in \cite{Ignat14_2} since negative values have been obtained for particle number density in the statistical theory at given assumption. However as later more in-depth researches have shown, this answer was obtained incorrectly. It turned to be that it is required to review a series of key points of the relativistic kinetic theory which depend on a sign of particle effective masses in order to strictly resolve the question of possibility to include negative effective masses into the theory. These key points appear at earlier stages of the relativistic kinetic theory preceding the definition of the macroscopic moments. The generalization of the theory to the case of multiple scalar fields is being considered in this article. When an attempt to make such a  generalization is made, it turns out that the developed theory contradicts to the fundamental principle of the action function additivity. The assumption of non-negativity of particles' effective mass is proved to be a theory statement which exactly leads to the contradiction. Thus, there raises a requirement to revise the kinetic theory of systems of scalar charged particles. The results of this revision are presented in the article.

\section{The Dynamics of Particles With The Scalar Interaction}
\subsection{The Canonical Equations of Motion}
A normalized invariant differential of volume of relativistic particle's $\Gamma$ 8-dimensional phase space,
being a vector bundle $\Gamma=X\times P$ %
with Riemann base $X(g)$ and vector bundle $P(X)$, with regard to a pair of canonically conjugated dynamic variables
 $x^{i} $ (configurational coordinates) and $P_{i}$ %
(coordinates of the generalized momentum)  is \cite{Bogolyub}\footnote{Here and further the universal system of units is used
 $G=c=\hbar =1$. In ordinary units it is $2\pi\to2\pi\hbar$.}:
\begin{eqnarray}\label{dG}
d\Gamma=\frac{\varrho}{(2\pi)^3}dXdP\equiv \nonumber\\
\frac{\varrho}{(2\pi)^3}dx^1dx^2dx^3dx^4dP_1dP_2dP_3dP_4,
\end{eqnarray}
where
\begin{eqnarray}
dX=\sqrt{-g}dx^1dx^2dx^3dx^4;\\
dP=\frac{1}{\sqrt{-g}}dP_1dP_2dP_3dP_4
\end{eqnarray}
are invariant differentials of volumes of the configurational and momentum spaces correspondingly and $\varrho$
is a degeneration factor; for particles with spin  $S$ it is $\varrho=2S+1$. %
Further to shorthand we introduce like phase space coordinates $\eta_a$, $a=\overline{1,8}$:
\begin{equation}
\eta_i\equiv x^i,\quad \eta_{i+4}=P_i,\quad (i=\overline{1,4}),
\end{equation}
where the expression for differential of volume of the phase space (\ref{dG}) takes the simplest form:
\begin{equation}\label{dG1}
d\Gamma=\frac{\varrho}{(2\pi)^3}\prod\limits_{a=1}^8 d\eta_a.
\end{equation}

The canonical equations of relativistic particle motion in the phase space $\Gamma$ have the following form (see e.g., \cite{Ignatev2}):

\begin{equation} \label{Eq1}
\frac{dx^{i} }{ds} =\frac{\partial H}{\partial P_{i} } ;\quad \quad \frac{dP_{i} }{ds} =-\frac{\partial H}{\partial x^{i} } ,
\end{equation}
 where $H(x,P)$ is a relativistically invariant Hamilton function and $u^i=dx^i/ds$ is a particle velocity vector.

In consequence of antisymmetry of the canonical equations of motion (\ref{Eq1}) and symmetry of the phase space
 (\ref{dG}) the differential relation \cite{Bogolyub} known as {\it Liouville theorem} in classical dynamics, is fulfilled
\begin{equation} \label{Liuvill}
\frac{d\Gamma}{ds}=0.
\end{equation}
According to this relation the phase volume of the world tube of particles is constant.

Calculating the full derivative of dynamic variables function $\Psi (x^{i} ,P_{k} )$ and taking into account (\ref{Eq1}) we obtain:
\begin{equation} \label{Eq2}
\frac{d\Psi }{ds} =[H,\Psi ],
\end{equation}
where the invariant Poisson brackets are introduced:
\begin{equation} \label{Eq3}
[H,\Psi ]=\frac{\partial H}{\partial P_{i}} \frac{\partial \Psi }{\partial x^{i} } -\frac{\partial H}{\partial x^{i} } \frac{\partial \Psi }{\partial P_{i} } \; .
\end{equation}
Let us note that Poisson bracket \eqref{Eq3} can be rewritten in the explicitly covariant form using
 {\it the operator of covariant Cartan differentiation}, %
$\widetilde{\nabla}_i$, \footnote{Covariant derivative in a bundle $\Gamma$ \cite{Cartan}.}, %
 (see e.g., \cite{Bogolyub})%
\footnote{Cartan covariant derivatives were first introduced into
relativistic statistics by A.A. Vlasov \cite{Vlasov}.}:
\begin{equation}\label{Cartan}%
\widetilde{\nabla}_i = \nabla_i +
\Gamma_{ij}^k P_k\frac{\partial}{\partial P_j},
\end{equation}
where $\nabla_i$ is an operator of covariant Ricci differentiation and $\Gamma^k_{ij}$ are Christoffel symbols of the second kind with
respect to metrics $g_{ij}$ of base $X$. %
Operator $\widetilde{\nabla}$ is defined in such a way that
\begin{equation}\label{9.11}%
\widetilde{\nabla}_iP_k \equiv 0
\end{equation}
and the following {\it symbolic} rule of functions differentiation is fulfilled:

\begin{equation}\label{9.13}%
\widetilde{\nabla}_i\Psi(x,P) = \nabla_i[\Psi(x],P).
\end{equation}
This rule means that in order to calculate Cartan derivative of function $\Psi(x,P)$ it would be enough to calculate its ordinary covariant
 derivative as if the momentum vector was covariant constant.
 Because of this equality the introduced operator is quite convenient
 for execution of differential and integral operations in the phase space %
$\Gamma$. Thus let us write the Poisson bracket \eqref{Eq3} in the explicitly covariant form:
\begin{equation}\label{H_Cart}
[H,\Psi ]\equiv \frac{\partial H}{\partial P_{i}} \widetilde{\nabla}_i\Psi-\frac{\partial \Psi}{\partial P_{i}} \widetilde{\nabla}_i H,
\end{equation}
Further, as a result of (\ref{Eq3}) the Hamilton function is an integral of particle motion:
\begin{equation} \label{Eq4}
\frac{dH}{ds} =[H,H]=0,\Rightarrow H= \Const.
\end{equation}
The relation (\ref{Eq4}) can be called a normalization ratio. Due to linearity of the Poisson bracket any continuously differentiable function
$f(H)$ is a Hamilton function as well. There is a sole possibility to introduce the invariant Hamilton function, square
by the particle generalized momentum at presence of gra\-vi\-ta\-tio\-nal and scalar fields. It is the following one:
\begin{equation} \label{Eq7}
H(x,P)=\frac{1}{2} \left[\psi(x)(P,P)-\varphi(x) \right],
\end{equation}
where $(a,b)$ here and further is a scalar product of the momentums $a$ and $b$ with respect to the base metrics:
\begin{equation}(a,b)=g_{ik} a^{i} b^{k} ,\nonumber\end{equation}
and $\psi(x)$ and $\varphi(x)$ are certain scalar functions of the scalar potentials.
Let us choose a zero nor\-ma\-li\-za\-ti\-on of the Hamilton function in the relation \eqref{Eq4} \cite{YuNewScalar1,Ignat_Popov}:
\begin{equation}\label{Eq7a}
H(x,P)=\frac{1}{2} \left[\psi(x)(P,P)-\varphi(x) \right]=0,
\end{equation}
where from we find:
\begin{equation}\label{Eq7b}
(P,P)=\frac{\varphi}{\psi};
\end{equation}
from the first group of the canonical equations of motion (\ref{Eq1}) %
we obtain the relation between the generalized momentum and particle velocity vector:
\begin{equation} \label{Eq10a}
u^{i} \equiv \frac{dx^{i} }{ds} =\psi P^{i} \Rightarrow P^{i} =\psi^{-1} u^{i} ,
\end{equation}
Substituting \eqref{Eq10a} into the normalization ratio (\ref{Eq7b}),
we obtain:
\begin{equation}(u,u)=\psi\varphi.\nonumber\end{equation}
Hence for particle velocity vector's normalization ratio fulfillment
\begin{equation} \label{Eq11}
(u,u)=1.
\end{equation}
it should be:
\noneq{\psi\varphi=1 \Rightarrow \psi=\varphi^{-1}.}
Thus particle's Hamilton function can be defined by a single scalar function $\varphi(x)$. Taking into account the last relation, let us write the Hamilton function in the final form:
\begin{equation}\label{Eq7 }
H(x,P)=\frac{1}{2} \left[\varphi^{-1}(x)(P,P)-\varphi(x) \right]=0.
\end{equation}
From the canonical equations (\ref{Eq1}) we can obtain the relation between
the generalized momentum and particle's velocity vector:
\begin{equation}
\label{Eq10} P^i=\varphi \frac{dx^i}{ds}.
\end{equation}
From (\ref{Eq7b}) it follows that the vector of generalized momentum if timelike:
\begin{equation}
\label{Eq8} (P,P)=\varphi^2\geqslant 0.
\end{equation}
Let us note a relation that could be useful further, being a consequence of  (\ref{H_Cart}), (\ref{Eq7}) and  (\ref{Eq8}):
\begin{equation} \label{Eq9}
[H,P^{k} ]=\nabla ^{k} \varphi \equiv g^{ik} \partial _{i} \varphi;
\end{equation}
where $\nabla^i\equiv g^{ik}\nabla_k$ is a covariant derivative's symbol.
\subsection{The Equations of Motion In The Lagrange Definition}

Let us obtain using  (\ref{Eq1}) the equations of motion in the Lagrange defintion \cite{Yubook1}:

\begin{equation} \label{Eq12}
\frac{d^{2} x^{i} }{ds^{2} } +\Gamma _{jk}^{i} \frac{dx^{j} }{ds} \frac{dx^{k} }{ds} =\partial _{,k} \ln |\varphi|{\rm {\mathcal P}}^{ik} ,
\end{equation}
 where:

\begin{equation} \label{Eq13}
{\rm {\mathcal P}}^{ik} ={\rm {\mathcal P}}^{ki} =g^{ik} -u^{i} u^{k}
\end{equation}
 is a tensor of orthogonal projection on the direction $u$ so that:

\begin{equation} \label{Eq14}
{\rm {\mathcal P}}^{ik} u_{k} \equiv 0;\quad {\rm {\mathcal P}}^{ik} g_{ik} \equiv 3.
\end{equation}
 The strict consequence of the velocity and acceleration vectors orthogonality follows from these relations and the Lagrange equations (\ref{Eq12}):

\begin{equation} \label{Eq15}
g_{ik} u^{i} \frac{du^{k} }{ds} \equiv 0.
\end{equation}
Let us notice that the Lagrange equations of motion (\ref{Eq12})
are invariant with respect to a sign of the scalar function $\varphi(x)$:
\begin{equation} \label{Eq16a}
\varphi(x)\rightarrow -\varphi(x).
\end{equation}
The Hamilton function (\ref{Eq7 }) at its zero normalization is also invariant to the transformation (\ref{Eq16a}). Therefore from (\ref{Eq8}), (\ref{Eq10}) and the Lagrange-Eiler equations (\ref{Eq12}) it follows that $\varphi $ scalar's square has a meaning of a square of \textit{ the effective inert mass of a particle, $m_{*} $ in the scalar field}:

\begin{equation} \label{Eq16}
\varphi^2 =m_{*}^2 .
\end{equation}
Let us notice that the following action function corresponds to the cited choice of the Hamilton function:
\begin{equation} \label{Eq17}
S=\int  m_{*} ds.
\end{equation}
It formally coincides with the Lagrange function of the relativistic particle with a rest mass $m_*$ in the gravitational field (see e.g., \cite{Land_Field}).

\subsection{The Choice of A Mass Function}
Let there be now  $n$ various scalar fields $\Phi_r$ in the system and each particle has %
$n$ corresponding fundamental scalar charges $q_r$ ($r=\overline{1,n}$) %
among which there can be some zero charges. Then the problem of choice of a function $m_{*} (\Phi_r)$ appears. %
Not specifying this function for now, let us note the next important circumstance: %
For equations of motion \eqref{Eq1} to allow the linear integral of motion %
$\Psi=(\xi,P)=\Const$, according to \eqref{Eq2} it is enough and sufficient that $[H,\Psi]=0$. This, in turn, is possible if and only if
 \cite{Ignatev14_1}:
\begin{equation}\label{Lin_Int}
(\xi,P)=\Const \Leftarrow\!\!\Rightarrow \ \Lee{\xi}\varphi g_{ik}=0,
\end{equation}
 where $\Lee{\xi}$ is a Lie derivative in the direction $\xi$ (see e.g., \cite{Petrov}).

 Let us consider statistical fields $g_{ik} $ and $\Phi_r$ allowing timelike Killing vector %
 $\xi ^{i} =\delta _{4}^{i} $ when the total energy of charged particle is conserved %
 $P_{4} =E_{0}=\Const >0$. Let us choose a system of coordinates where $g_{\alpha 4} =0, \alpha,\beta=\overline{1,3}$ so that
  $x^{4} $ coincides with the world time $t$. Then from the relations connecting the kinematic velocity vector
   $u^{i} $ with the total momentum of a particle $P_{i} $ \eqref{Eq10} it follows:
\begin{equation}
P_{4} ds=\varphi dt.\nonumber
\end{equation}
Therefore, if we {\it require} to conserve the same orientation of the world time $t$ and proper time $s$ (i.e. $u^4=dt/ds>0$) then the mass function should be chosen in such a way that it is always nonnegative:

\begin{equation}
m_{*} =|\varphi| >0.\nonumber
\end{equation}
However such a conservative approach had been being used in Author's papers
\cite{Ignatev1} -- \cite{Ignatev4} and seeming at first sight the correct one, turns to be contradicting
to the more fundamental principle of the Lagrange function's additivity. As it was shown in \cite{Ignatev14_1}, %
the negativeness of the particle's effective mass function
does not lead to any contradictions at the level of microscopic dynamics since the observable momentum of particle
(as well as the 3-dimensional velocity $v^\alpha=u^\alpha/u^4$)
conserves its orientation as opposed to the unobservable kinematic 4-velocity of a particle $u^i$:
\begin{equation}\label{Pp}
p^i=m_*\frac{dx^i}{ds}\equiv P^i.
\end{equation}
Then the choice of a linear mass function in \eqref{Eq17} corresponds to
the principle of action function's additivity:
\begin{equation} \label{Eq27}
m_*=m_0+\sum\limits_r  q^{(r)}\Phi_r,
\end{equation}
where $m_0$ is a certain initial rest mass and $q^{(r)}$ is
a charge of particle relative to the scalar field $\Phi_r$ which we presume functional independent.
At $m_0=0$ the entire rest mass of particles is ensured by interaction with scalar fields:
\begin{equation}\label{Eq28}
m_*=\varphi=\sum\limits_r q_r\Phi_r.
\end{equation}
This choice satisfies also aesthetic criteria since in this case
the Hamilton function%
\eqref{Eq7} does not depend on the rest mass. From the other hand it is clear that if choose function %
$\varphi (\Phi)$ in form \eqref{Eq27} the Lagrange equations \eqref{Eq12} become symmetric with respect to the change %
 $\Phi_r \to -\Phi_r $ or $q_r \to -q_r $. Therefore if particles $a$ and %
antiparticles $\bar{a}$ are different in sign of scalar charges then $\bar{m}_*=-m_*$, but trajectories of
particle and antiparticle in gravitational and scalar fields are identical. In such case according to  %
\eqref{Eq10a} 4-vectors of their kinematic velocity are different in sign: $\bar{u}^i=-u^i$, and %
have coinciding vectors of the observable generalized momentum $\bar{P}^i=P^i$. Let us note that as this takes place, vectors %
of the 3-dimensional velocity $v^\alpha=u^\alpha/u^4$ ($\alpha=\overline{1,3}$) for particles and antiparticles also coincide with each other: $\bar{v}^\alpha=v^\alpha$. Nevertheless we do not assume $m_0=0$ {\it for now} postponing this for the future.

Having chosen the effective mass in form (\ref{Eq27}), the Hamilton function \eqref{Eq7 }
and the normalization ratio (\ref{Eq8}) for the generalized momentum take form:
\begin{equation}\label{H,m}
H(x,P)=\frac{1}{2} \left[m_*^{-1}(x)(P,P)-m_*\right]=0,
\end{equation}
\begin{equation}\label{P_norm}
(P,P)=m^2_*.
\end{equation}
Let us notice identities valid for the Hamilton function \eqref{H,m} which could be useful in future:
\begin{equation}\label{nabla_H}
\widetilde{\nabla}_iH=-\nabla_i m_*,
\end{equation}
\begin{equation}\label{HPsi}
[H,\Psi]=\frac{1}{m_*}P^i\widetilde{\nabla}_i\Psi+\partial_i m_*\frac{\partial \Psi}{\partial P_i},
\end{equation}
where $\Psi(x,P)$ is an arbitrary function.

\subsection{The Quantum Equations}
From the classical Hamilton function \eqref{H,m} with a help of standard substitution
:\footnote{In this section we temporarily renounce the use of the universal system of units where $\hbar=1$.}
\begin{equation}\label{quant_trans}
P_i\rightarrow i\hbar\nabla_i
\end{equation}
we obtain the Hamilton operator:
\begin{equation}\label{Hamiltonian}
\hat{{\rm H}}=-\frac{1}{2}m_*^{-1}(\hbar^2 g^{ik}\nabla_i\nabla_k + m^2_*).
\end{equation}
Thus for a free massive scalar field we obtain wave equations in form of standard Klein-Gordon equations with
the only difference being bosons rest mass changed to the effective mass \cite{Ignatev14_1}:
\begin{equation}\label{free_bozon}
(\square+m_*^2/\hbar^2)\Psi=0.
\end{equation}
For free fermions we obtain corresponding Dirac equations:
\begin{equation}\label{Dirac_eq}
(\hbar\gamma^i\nabla_i+m_*)\Psi=0,
\end{equation}
where $\gamma$ are spinors.

Let us notice that from (\ref{free_bozon}) at substitution $\Psi=\Phi$ and choice of the mass function $m_*=q\Phi$
as well as for the mass function $m=|q\Phi|$ it straight away follows the equation
of the scalar field with cubic nonlinearity:
\begin{equation}\label{Phi3}
\square\Phi+(q^2/\hbar^2)\Phi^3=0.
\end{equation}
Thus the self-action constant in the equation of the scalar field
takes well-defined meaning:
\noneq{\lambda=\frac{q^2}{\hbar^2},}
i.e., it is defined identically to the fine structure constant for the electromagnetic field.

\section{Calculation of the macroscopic means}
In order to extend the general relativistic kinetic theory to the case of negative effective masses of elementary particles it is required to review the key points of the relativistic kinetic theory which could depend on a sign of particle mass. It has been revealed that there are only two of such key points and they are defined by two circumstances: the relation between proper microscopic time of a particle and  proper time of macroscopic observers and the relation between particle velocity vector and its generalized momentum.

\subsection{The Invariant Distribution Function}
Generalized formalism of invariant distribution functions was developed in articles \cite{Yubook1,Bogolyub}. It is required
to thoroughly apply this formalism to the case being considered in order to account the possibility of the negative sign
 of the effective mass of particles.
According to this formalism, to define the macroscopic means
in the relativistic phase space
it is required to define the unit timelike field of the macroscopic observers $U_i(x) :(U,U)=1$. The synchronization
of measurement acts for individual particles is carried out by the clocks of such observers. This timelike field in its turn defines
certain spacelike three-diemensional surface $V_3$, displacements along which  $\delta x^i$ are orthogonal to this field:
\begin{equation}\label{Sigma}
V_3:\; \delta x^iU_i=0,
\end{equation}
and displacements along this field $dx^i$ define the synchronized proper time $\tau$ of observers:
\begin{equation}\label{tau}
\frac{dx^i}{d\tau}=U^i \Leftrightarrow \frac{dx^i}{d\tau}U_i=1\Rightarrow d\tau=dx^iU_i.
\end{equation}
Thus in the observers' macroscopic reference frame:
\begin{equation}\label{dXt}
X=V\times T \Rightarrow dX=dVd\tau.
\end{equation}

Apparently, the connection between the proper time of particle $s$ and
the synchronized proper time $\tau$ of observers in each point of the configurational space is established by the relation:%
\begin{equation}\label{dtauds}
\frac{d\tau}{ds}=U_i\frac{dx^i}{ds}.
\end{equation}
Now taking into account relation (\ref{Pp}) being the consequence of the canonical equations (\ref{Eq1}), we finally obtain:
\begin{equation}\label{dsdtau}
\frac{d\tau}{ds}=m_*^{-1}(U,P)\Rightarrow \frac{1}{m_*(s)}\frac{ds}{d\tau}=\frac{1}{(U(\tau),P(s))}.
\end{equation}
The relation (\ref{s(tau)}) can be considered as a differential equation in the
function $s(\tau)$, resolving which we define the relation between
 proper time of particle and time
measured by the clocks of
synchronized observers:
\begin{equation}\label{s(tau)}
s=s(\tau).
\end{equation}

The invariant 8-dimensional identical particles distribution function $F(x,P)$ is introduced in the following way \cite{Yubook1,Bogolyub}.
Let phase trajectory of particle being defined by the canonical equations (\ref{Eq1}) in the phase space $\Gamma=P(X)\times X$ is:
\begin{equation}\label{x(s),P(s)}
x^i=x^i(s);\quad P_i=P_i(s)\Rightarrow \eta_a=\eta_a(s)\; (a=\overline{1,8}),
\end{equation}
where $s$ is a proper time of particle.
Then the number of particles registered by the observers in range $d\Gamma$ of the phase space can be defined as \cite{Yubook1,Bogolyub}:
\begin{equation}\label{dN}
dN(\tau)=F(x,P)\delta(s-s(\tau))d\Gamma.
\end{equation}
Let us notice that particle number is a scalar depending, however, on the choisce of observers' field $U^i$, i.e., on the choice of reference frame in Riemann space $X$,
while 8-dimensional distribution function itself $F(x,P)$ being introduced by the relation (\ref{dN}), is an invariant in phase space $\Gamma$. Let us note also that it is impossible to provide any other definition for the invariant distribution function in the 8-dimensional phase space. All definitions of this function which were introduced earlier are special cases (\ref{dN}) realized in the dedicated reference frames.

\subsection{The Macroscopic Means of Dynamic Functions}
The definition of the invariant distribution function (\ref{dN}) and other dynamic functions together with it is a first key point of the relativistic kinetic theory which depends on the sign of particles' mass.
Let $\psi(x,P)\equiv\psi(\eta)$ is a certain scalar function of dynamic variables. %
Then according to (\ref{dN}) its macroscopic mean $\Psi(\tau)$ in range $\Omega \subset\Gamma$ %
is defined in the following way:
\begin{eqnarray}\label{Psi(tau)dG}
\Psi(\tau)=\int\limits_\Omega \psi(\eta(s))dN=\nonumber\\
\int\limits_\Omega F(\eta(s))\psi(\eta(s))\delta(s-s(\tau))d\Gamma.
\end{eqnarray}
Let us emphasize one more time that distribution function $F(x,P)$ is an invariant in the phase space as opposed to the macroscopic means which are
always defined with respect to observers' field $U$. Putting further in accordance with
 \eqref{Sigma}, \eqref{tau} $X=V\times T$ $\Rightarrow dX=dVdt$, let us write expression запишем выражение %
(\ref{Psi(tau)dG}) in the explicit form:
\begin{eqnarray}\label{Psi(tau)dVt}
\Psi(\tau)=\frac{2S+1}{(2\pi)^3}\int\limits_V dV\int\limits_T dt\times\nonumber\\
\int\limits_{P(X)} dP \psi(\eta) F(\eta)\psi(\eta)\delta(s-s(\tau))
\end{eqnarray}
To carry out integration over $t$ in (\ref{Psi(tau)dVt}) let us take into account the relation between $t(s)$ (\ref{Pp}) and %
 $\tau(s)$ (\ref{s(tau)}) and the property of Dirac $\delta$-function:
\begin{eqnarray}\label{int_s}
\delta(s-s(\tau))dt=\left|\frac{d\tau}{ds}\right|\delta(t-\tau)ds\nonumber\\
\equiv |m_*|^{-1}(U,P)\delta(t-\tau)dt.
\end{eqnarray}
At derivation of (\ref{int_s}) we took into account the fact that
generalized momentum's orientation
does not depend on
the sign of the
effective mass
as opposed to kinematic velocity vector's orientation.
Taking now the account of (\ref{int_s}) and (\ref{Psi(tau)dVt}) and conducting integration over time coordinate, we finally obtain:
\begin{eqnarray}\label{Psi_7}
\Psi(\tau)=\frac{2S+1}{(2\pi)^3}\int\limits_V  U_i dV \frac{1}{|m_*|}\times \nonumber\\
\int\limits_{P(X)} P^i dP \psi(\eta)F(\eta)\psi(\eta)
\end{eqnarray}
Particularly, supposing $\psi=1$, we obtain the total number of particles in range $V$:
\begin{eqnarray}\label{N}
N(V)=\frac{2S+1}{(2\pi)^3}\int\limits_V  U_i dV \frac{1}{|m_*|}\times\nonumber\\
\int\limits_{P(X)} P^i F(\eta)dP\equiv \int\limits_V  (U,n) dV,
\end{eqnarray}
where  {\it particle flux density vector} is introduced:
\begin{equation}\label{ni}
n^i(x)=\frac{2S+1}{(2\pi)^3|m_*|}\int\limits_{P(X)} P^i F(\eta)dP
\end{equation}
This factor of effective mass modulus was not taken into account in the previous articles because of above cited conservatism reason.

\subsection{Transformation to the 7-Dimensional Distribution Function}
As a result of generalized momentum normalization ratio (\ref{P_norm}) the invariant 8-dimensional distribution function
 $F(x,P)$ is a singular one on the mass surface. Therefore it is necessary to
 introduce a distribution function non-singular on this surface $f(x,P)$ by means of the relation:
\begin{equation}\label{f}
F(x,P)=\delta(H(x,P))f(x,P).
\end{equation}
Calculating the relation
\begin{equation}
F(x,P)dP\equiv\!\!\! \frac{1}{\sqrt{-g}}dP_1dP_2dP_3dP_4 \delta(H(x,P))f(x,P)\nonumber
\end{equation}
with a help of Dirac $\delta$-function's properties and the explicit form of the Hamilton function \eqref{H,m}, we find:
\begin{eqnarray}\label{F,f}
F(x,P)dP=\hskip 3cm \nonumber\\
\frac{1}{\sqrt{-g}}dP_1dP_2dP_3  \frac{|m_*|}{P^4_+}\delta(P_4-P_4^+)f(x,P)\times\nonumber\\
\equiv |m_*|dP_0 \delta(P_4-P_4^+)f(x,P),
\end{eqnarray}
where $P_4^+\equiv P_4$ is a positive root of the normalization equation (\ref{P_norm})
and
\begin{equation}\label{dP_0}
dP_0=\frac{1}{\sqrt{-g}}\frac{dP_1dP_2dP_3}{P^4}
\end{equation}
is a differential of volume of the 3-dimensional momentum space.

As a result formulas for the macroscopic means \eqref{Psi_7} can be written through the  %
{\it 7-dimensional distribution function} $f(x,P)$ in the following form:
\begin{eqnarray}\label{Psi_6}
\Psi(\tau)=\frac{2S+1}{(2\pi)^3}\int\limits_V  U_i dV\!\!\!
 \int\limits_{P_+(X)}\!\! P^i dP_0 \psi(\eta)f(\eta),
\end{eqnarray}
where it is necessary to substitute $P_4$ by the positive root of the mass surface equation and $P_+$
is an upper part of the mass surface pseudosphere. Thus at transformation to the 7-dimensional distribution function the explicit
dependency on the effective mass disappears.

Therefore the next symbolic rule being understood in a sense of
integration over corresponding phase
volumes, is valid:
\begin{eqnarray}\label{psi_rule}
\psi(\eta) F(\eta)\delta(s-s(\tau))d\Gamma\rightarrow \nonumber\\
\psi(\tilde{\eta})f(\tilde{\eta})(U,P)dVdP_0,
\end{eqnarray}
where $\tilde{\eta}$ are dynamic variables on the 6-dimensional subspace $\Gamma_0(\tau)=V\times P_0\subset \Gamma$
with the differential of volume:
\begin{equation}\label{G0}
d\Gamma_0=dVdP_0.
\end{equation}
In particular, for the {\it particle number flux density vector}%
\footnote{Согласно J. Synge \cite{Sing}.} \eqref{ni} %
from \eqref{Psi_6} we find:
\begin{equation}\label{ni}
n^i(x)=\frac{2S+1}{(2\pi)^3}\int\limits_{P_0(X)} P^i f(\eta)dP.
\end{equation}

Let us notice that in \cite{Ignat14_2} the incorrect conclusion was made regarding the absence of dependency between
particle number flux density vector and the effective mass sign. This conclusion was a result of the fact that macroscopic means' calculation
has been started with a calculation of distribution function moments while in case of negative masses the correct calculations
should have been started on the earlier stage, namely they should have been started with integral relations of form \eqref{Psi(tau)dG}.


\end{document}